\documentstyle[preprint,aps]{revtex}
\tightenlines

\begin{document}

\title{Implications of Cosmic Repulsion for Gravitational 
Theory\footnote{astro-ph/9804335, 
April 30, 1998}}

\author{\normalsize{Philip D. Mannheim} \\
\normalsize{Department of Physics,
University of Connecticut, Storrs, CT 06269} \\ 
\normalsize{mannheim@uconnvm.uconn.edu} \\}

\maketitle

\begin{abstract}
In this paper we present a general, model independent analysis
of a recently detected apparent cosmic repulsion, and discuss its 
potential implications for gravitational theory. In particular, we 
show that a negatively spatially curved universe acts like a 
diverging refractive medium, to thus naturally cause galaxies to 
accelerate away from each other. Additionally, we show that it is 
possible for a cosmic acceleration to only be temporary, with 
some accelerating universes actually being able to subsequently 
recontract.
\end{abstract}
\bigskip 

In cosmology the assumption of 3 space isotropy and homogeneity 
forces the geometric line element to assume the Robertson Walker 
(RW) form 
\begin{equation}
d\tau^2 =c^2dt^2-R^2(t)[(1-kr^2)^{-1}dr^2+r^2d\Omega]
\label{1}
\end{equation}
with cosmology thus being defined by two quantities, the 
scale factor $R(t)$ and the spatial curvature $k$. As such these 
quantities are in principle independent, though they can become
related upon the imposition of a specific dynamical cosmological 
model. To distinguish candidate cosmologies it is thus desirable to 
try to extract out information about these two quantities from 
observational data in as model independent a way as possible. 
Standing in the way of doing this in the past has been the fact 
that while the current value $H(t_0)=\dot{R}(t_0)/R(t_0)$ of the 
Hubble parameter can be determined by measurements of the nearby 
universe, a determination of its derivative, viz. the deceleration 
parameter $q(t)=-\ddot{R}(t)R(t)/\dot{R}^2(t)$, requires 
measurements over cosmologically separated time scales, 
measurements that have only very recently become available through 
the use of type Ia supernovae standard candles at high 
$z$ redshift. While the newly reported Hubble plot data 
\cite{Garnavich1998,Perlmutter1998} do not permit 
the extraction of definitive values for the RW cosmological 
parameters, nonetheless they do sharply reduce the space available 
to these parameters in a way which has turned out to be quite 
surprising as it revealed the apparent presence of a cosmic 
repulsion, a repulsion which had not at all been expected in the 
standard big bang Einstein-Friedmann cosmology. In a recent paper 
\cite{Mannheim1998} it was shown that negative spatial curvature 
$k$ could actually account for this cosmic repulsion, and that even 
while such non-zero curvature is not anticipated in the standard 
cosmology, nonetheless another cosmology was presented, viz. 
conformal cosmology, in which $k$ could naturally (i.e. without 
fine tuning) actually be negative. (That $k$ would be negative in
conformal cosmology and that $q(t_0)$ would likewise be 
substantially smaller than its canonically expected value 
\cite{Guth1981} of $q(t_0)=1/2$ had already been noted by Mannheim 
quite some time ago 
\cite{Mannheim1992,Mannheim1995,Mannheim1997,Mannheim1996}). 
In this paper we explore some general, model independent aspects of 
negative $k$ cosmologies and provide an interpretation of the 
cosmic repulsion they generate as being characteristic of a spatial 
geometry which acts like a diverging refractive medium. 

While the intent of this paper is to  discuss some kinematic 
aspects of curvature, it is nonetheless useful for both orientation 
purposes and for identifying the key features of the new data to 
first consider the implications of the standard 
Einstein-Friedmann dynamical evolution equation, viz.
\begin{equation}
\dot{R}^2(t) +kc^2=\dot{R}^2(t)\Omega_{M}(t)
+\dot{R}^2(t)\Omega_{V}(t)
\label{2}
\end{equation}
where $\Omega_{M}(t)=8\pi G\rho_{M}(t)/3c^2H^2(t)$ is due to 
ordinary matter and $\Omega_{V}(t)=8\pi G\Lambda/3cH^2(t)$ is
due to a cosmological constant $\Lambda$. For such an evolution
equation, in the case where the energy density takes the form 
$\rho_{M}(t)=A/R^n(t)$, the deceleration parameter can be written
in the convenient form
\begin{equation}
q(t)=(n/2-1)\Omega_{M}(t)-\Omega_{V}(t)=
(n/2-1)(1+kc^2/\dot{R}^2(t)) - n\Omega_{V}(t)/2 
\label{3}
\end{equation}
\noindent
We thus see that both negative $k$ and positive $\Omega_{V}(t)$ 
serve to reduce $q(t)$ below its pure matter ($\Omega_{M}(t)=1$, 
$k=0$, $\Omega_{V}(t)=0$) flat inflationary universe value of 
$n/2-1$, while positive $k$ and negative $\Omega_{V}(t)$ serve to 
increase $q(t)$. In order to understand why non-zero $k$ modifies 
the deceleration parameter, we recall that the standard wisdom 
regarding the big bang is that after an initial explosion the
mutual gravitational attraction of the galaxies would then serve to 
brake the expansion so that the universe would gradually slow down.
However, that discussion presupposes that the galaxies are 
propagating in an inert, empty space (viz. $k=0$) which itself has 
no dynamical consequences. However, once $k$ is non-zero, the 
galaxies then propagate in a non-trivial geometric medium, a medium 
which then can explicitly participate dynamically, with the 
gravitational field itself then present in the medium then being 
able to accelerate or decelerate the galaxies according to the sign 
of its spatial curvature. While the observational detection of a 
cosmic repulsion can thus be thought of as a measurement of the 
global topology of the universe, this conclusion is not yet 
warranted (it will potentially become warranted below when we 
consider conformal cosmology), since the above standard cosmology 
evolution equation is not thought to be reliable for $k \neq 0$ 
topologies where it would then possess a severe fine-tuning 
flatness problem.
 
Given the issue of the reliability of Eq. (\ref{2}) away from 
$k=0$, it would be instructive to see if we could understand how 
negative spatial curvature does serve to reduce the magnitude of 
the deceleration parameter without needing to appeal to any 
cosmological evolution equation at all. To this end we first 
consider the purely kinematic lightlike geodesics associated with 
the RW line element, with the source coordinates $(t_1,r_1)$ of 
some emitter of light (at luminosity distance 
$d_L=r_1R^2(t_0)/R(t_1)$ and redshift $z=R(t_0)/R(t_1)-1$) being 
related to those of an observer at $(t_0,0)$ via
\begin{equation}
p(t_0)-p(t_1)=\int^{t_0}_{t_1} dt cR(t)^{-1}=
\int_{0}^{r_1} dr (1-kr^2)^{-1/2}
\label{3a}
\end{equation}
where the 
convenient $p(t)=\int^t dt cR(t)^{-1}$ is the conformal time. Since
the quantity $\int_{0}^{r_1} dr (1-kr^2)^{-1/2}$ is less than $r_1$
if $k$ is negative, the light signal effectively arrives at the
source at $r=0$ in less (conformal) time in a $k<0$ space than it 
would have needed to travel from the same originating point $r_1$ 
had the intervening space been flat. The negative curvature of the 
space thus effectively gives the light a push (positive curvature 
would give a pull) to enable light to 
transport energy faster than it would otherwise be able to in empty 
space, with negative curvature thus acting like a diverging
refractive medium. Such a medium would then serve to cause light to 
diverge away from a source, so that, analogously, galaxies 
propagating in such a medium would then be propelled away 
from each other.

Further insight into the refractive nature of $k\neq 0$ spaces can 
be obtained by considering the Maxwell equations 
$F^{\mu \nu}_{\phantom{\mu \nu};\nu}=0$, 
$F_{\mu \nu ; \lambda}+ F_{\lambda \mu ; \nu}+ 
F_{\nu \lambda ; \mu }=0$ in background RW 
spacetimes. These equations simplify a great deal if the RW line 
element is written in isotropic coordinates
\begin{equation}
d\tau^2 =R^2(p)(dp^2-\gamma_{ij}dx^idx^j)
\label{4}
\end{equation}
(where $\gamma_{ij}=f^2(\rho)\delta_{ij}$,
$\rho^2=(x^1)^2+(x^2)^2+(x^3)^2$, and 
$f(\rho)=(1+k\rho^2/4)^{-1}$), and 
are found \cite{Plebanski1960,Mashoon1973,DengandMannheim1988} 
to take the form
\begin{eqnarray}
{\bf \nabla}\cdot (f(\rho){\bf E})=0~,~
{\bf \nabla}\times {\bf E}=-f(\rho)\partial {\bf H}/\partial p~,
\nonumber \\
{\bf \nabla}\cdot (f(\rho){\bf H})=0~,~
{\bf \nabla}\times {\bf H}=f(\rho)\partial {\bf E}/\partial p~~
\label{5}
\end{eqnarray}
where we have introduced the fields $F_{0i}=-E_i$, 
$F_{ij}=\epsilon_{ijk}H_k$. On defining ${\bf D}=\epsilon {\bf E}$, 
${\bf B}=\mu {\bf H}$ we immediately recognize Eqs. (\ref{5}) as 
being precisely of the same form as that of the (flat spacetime) 
Maxwell equations in a constitutive medium with $\epsilon=f(\rho)$, 
$\mu=f(\rho)$, so that curvature is indeed seen to act just like a 
refractive medium. To explore the explicit propagation 
properties of light in this medium, we note that even though these 
curved space Maxwell equations mix the various components of 
${\bf E}$ and ${\bf H}$, it turns out \cite{MannheimandKazanas1988} 
that these equations are actually diagonal in the radial components 
of the electromagnetic field and explicitly yield (with $S(p,x^i)$ 
denoting either $\rho E_{\rho}$ or $\rho H_{\rho}$)
\begin{equation}
(\partial^2/\partial p^2 +k -\Delta^{(3)})S(p,x^i)=0
\label{6}
\end{equation}
as the relevant wave equation of interest. 
The covariant 3 space d'Alambertian $\Delta^{(3)}=
(\gamma)^{-1/2}\partial_i (\gamma)^{1/2}\gamma^{ij}\partial_j$ 
admits of a separation constant $-\lambda^2$,
so that the frequencies of this radial equation then obey
$\omega^2/c^2=\lambda^2+k$, with the explicit wave functions being
tabulated in Refs. \cite{MannheimandKazanas1988} and 
\cite{DengandMannheim1988}. Spatial curvature is thus seen to 
act as a tachyonic mass when $k<0$, to thus effectively give faster 
than light propagation. For this radial wave equation an effective 
frequency dependent refractive index can be defined as 
$n(\omega)=c\lambda/\omega=(1-kc^2/\omega^2)^{1/2}$, with the group 
velocity of the associated waves then being given by 
$v_{g}=d\omega/d\lambda=
c(1-kc^2/\omega^2)^{1/2}$. A negative curvature space is thus 
seen to act like a diverging dispersive medium, with the group 
velocity of light in the medium being explicitly greater than one 
when $k$ is negative. Moreover, since 
\cite{MannheimandKazanas1988} Eq. (\ref{6}) can also be written in 
the form $S^{;\mu}_{\phantom{;\mu};\mu}=0$, 
we recognize it to be the fully covariant wave equation associated 
with the propagation of a scalar field in the background RW 
geometry of Eq. (\ref{1}). We thus see that our discussion 
holds for particles other than photons as well. Moreover, if we 
set $S(x)=exp(iP(x))$, the eikonal
phase $P(x)$ is then found to obey the equation  
$P^{;\mu}P_{;\mu}-iP^{;\mu}_{\phantom{;\mu};\mu}=0$, a condition
which reduces to $P^{;\mu}P_{;\mu}=0$ in the short wavelength 
limit. From the associated condition $P^{;\mu}P_{;\mu ;\nu}=0$ it 
then follows that $P^{;\mu}P_{;\nu ;\mu}=0$. Since normals
to the wavefronts obey the eikonal relation $P^{;\mu}=dx^{\mu}/dq$
where $q$ is a convenient parameter which measures distance along
the normals, we thus see \cite{Mannheim1993} that in the eikonal 
approximation the rays associated with Eq. (\ref{6}) precisely 
follow the RW geodesics of Eq. (\ref{3a}). Moreover, exactly the 
same analysis holds \cite{Mannheim1993} for the massive wave 
equation $S^{;\mu}_{\phantom{;\mu};\mu}-m^2S=0$ as well. Thus both 
massless and massive test particles will be accelerated by 
background negative spatial curvature no matter what the form of 
the gravitational equations of motion.
 
Now while negative curvature does kinematically give repulsion, 
whether or not this effect dominates does depend on the 
cosmological evolution equations, and to explore this issue we 
next need to consider the effect of the presence of a possible 
cosmological constant. To this end it is possible to
obtain some general information directly from the symmetry of 
the geometry without actually needing to explicitly 
assume the validity of the Einstein equations at 
all.\footnote{Indeed, it is the symmetry of the geometry 
rather than the details of the dynamics which usually enables us 
to find exact general relativistic solutions in the first place.} 
Specifically, let us assume only that there is a cosmology which is 
described by an underlying maximally 4 space symmetric de Sitter 
(or anti de Sitter) geometry with 4 space curvature $\alpha$, 
viz. a cosmology in which the Riemann tensor takes the form
\begin{equation}
R^{\lambda\rho\sigma\nu} =\alpha(g^{\sigma \rho}g^{\lambda \nu}
-g^{\nu \rho}g^{\lambda\sigma}) 
\label{7}
\end{equation}
\noindent
In such a geometry the contraction of Eq. (\ref{7}) leads to
\begin{equation}
R^{\mu \nu} - g^{\mu \nu} R^{\sigma}_{\phantom {\sigma} \sigma}/2
= 3\alpha g^{\mu \nu}
\label{8}
\end{equation}
a purely geometric relation which will hold in any cosmology for
which de Sitter geometry is an allowed solution, a relation whose 
validity in no way requires the a priori assumption of the Einstein 
equations, equations which would have to hold as equations of 
motion for all possible geometries and not merely be a property of 
some particular solution to them (Eq. (\ref{8}) also holds in the 
de Sitter solution to conformal cosmology - though not of course in 
other solutions to conformal gravity).\footnote{In passing we note 
the one way nature of the above discussion. While Eq. (\ref{8}) 
follows from Eq. (\ref{7}) the reverse is not true, since Eq. 
(\ref{8}) could in principle be satisfied by geometries with 
symmetry lower than the maximally 4 space symmetric one of 
Eq. (\ref{7}).} Rather than solve  Eq. (\ref{8}) directly in a de 
Sitter space, it is more convenient for our purposes here to 
reexpress Eq. (\ref{8}) in RW coordinates, so that Eq. (\ref{8}) 
then takes the convenient form
\begin{equation}
\dot{R}^2(t) +kc^2=\alpha c^2 R^2(t)
\label{9}
\end{equation}
\noindent
It is important to realize we are still in a de Sitter geometry
since the Riemann tensor of the RW geometry of Eq. (\ref{1}) is 
explicitly found to reduce to that of Eq. (\ref{7}) when 
Eq. (\ref{8}) is imposed (for instance 
$R_{1212}= -R^2r^2(\dot{R}^2 +kc^2)/c^2(1-kr^2)$
and $R_{1010}=\ddot{R}R/c^2(1-kr^2)$
immediately become the $R_{1212}$ and $R_{1010}$ components of 
Eq. (\ref{7}) when it is written in these same RW coordinates.) 
With this writing of the de Sitter geometry in RW coordinates we 
are then able to relate the 3 and 4 space topologies and 
curvatures, something we investigate below. 

Solutions to Eq. (\ref{9}) are readily obtained  
once the signs of $k$ and $\alpha$ are specified. For $\alpha$ 
greater than zero solutions can be found for all three possible 
choices for $k$, and they are found to take the form
\begin{eqnarray}
R(t,\alpha>0,k>0)=(k/\alpha)^{1/2}cosh(\alpha^{1/2}ct)
\nonumber \\
R(t,\alpha>0,k=0)=R(t=0)exp(\alpha^{1/2}ct)
\nonumber \\
R(t,\alpha>0,k<0)=(-k/\alpha)^{1/2}sinh(\alpha^{1/2}ct)
\label{10}
\end{eqnarray}
whereas for $\alpha\leq 0$ the only allowed topology is $k<0$, with 
the solutions then taking the form
\begin{eqnarray}
R(t,\alpha=0,k<0)=(-k)^{1/2}ct
\nonumber \\
R(t,\alpha<0,k<0)=(k/\alpha)^{1/2}sin((-\alpha)^{1/2}ct)
\label{11}
\end{eqnarray}
\noindent
In these solutions the deceleration parameter takes the 
respective forms
\begin{eqnarray}
q(t,\alpha>0,k>0)=-coth^2(\alpha^{1/2}ct)
\nonumber \\
q(t,\alpha>0,k=0)=-1
\nonumber \\
q(t,\alpha>0,k<0)=-tanh^2(\alpha^{1/2}ct)
\nonumber \\
q(t,\alpha=0,k<0)=0
\nonumber \\
q(t,\alpha<0,k<0)=tan^2((-\alpha)^{1/2}ct)
\label{12}
\end{eqnarray}
\noindent
As we see, all the $\alpha \geq 0$ de Sitter cosmologies expand for 
ever, while only the  $\alpha <0$ anti de Sitter one is able to 
expand and then recontract. Additionally, all of the $\alpha >0$ 
cosmologies have negative deceleration parameters, and they thus  
accelerate indefinitely (such acceleration is of course the 
reason why inflation inflates) and expand indefinitely. However, 
the  $\alpha<0$ cosmology has a positive deceleration parameter,
a deceleration which is large enough to actually stop and reverse 
the expansion altogether (in this case the acceleration obeys the 
simple harmonic oscillator relation $\ddot{R}(t) 
=c^2\alpha R(t)$). The $\alpha=0$ cosmology is just at the 
$q(t)=0$ borderline between acceleration and deceleration, and
corresponds to the $k$ dominated ever expanding $R(t)=(-k)^{1/2}ct$ 
cosmology. The correlation between the sign of $\alpha$ and the
sign of $q(t)$ precisely mirrors our earlier discussion of the  
implications of the Einstein evolution equation of Eq. 
(\ref{3}), to thus not only explain the connection between
the signs of $q(t)$ and $\Omega_{V}(t)$ exhibited in Eq. (\ref{3}), 
but to also show that the correlation between $q(t)$ and $\alpha$ 
is actually more general than that, as it holds independent of the 
assumption of any particular set of gravitational equations of 
motion.   

The appearance of the three different $\alpha>0$ solutions given 
above is at first sight puzzling since there is only one $\alpha>0$
de Sitter geometry. However, our discussion above only involved
the local equivalence of the de Sitter and RW Riemann tensors, and
did not address the global topologies of those geometries. Thus
while the topologically open anti de Sitter $\alpha<0$ geometry is 
both locally and globally equivalent to a topologically open $k<0$ 
RW geometry, the situation is somewhat different for the 
$\alpha>0$ case. Since the de Sitter geometry is topologically
closed it corresponds both locally and globally to the 
topologically closed $k>0$ RW geometry, so these two geometries do 
coincide in the $\alpha>0$, $k>0$ case. However, the two other 
$\alpha >0$ solutions are not topologically closed. They thus each  
correspond to only one coordinate patch of a de Sitter space, and
are thus topologically different from a de Sitter geometry, to
thus represent RW solutions which are distinct from de Sitter ones.
While this distinction may only appear to be geometrical, its 
significance (as will become explicitly apparent in our study of 
conformal cosmology below) lies in the fact that as soon as a 
perfect fluid of matter fields is introduced into the cosmology, 
the symmetry is immediately lowered from maximally 4 space to 
maximally 3 space symmetric. While such cosmologies could then have 
an evolution in which $R(t)$ might actually look like one of the 
solutions of Eq. (\ref{10}) in some epoch, nonetheless the global 
topology associated with the presence of the fluid would remain 
fixed once and for all and would not itself evolve. 

Having now identified the kinematic cosmological aspects of both 
non-trivial spatial RW curvature and of any underlying de Sitter 
or anti de Sitter geometry, we turn now to a discussion of the 
dynamical implications of the new supernovae 
$d_L$ versus $z$ Hubble plot data themselves, data 
whose great utility is that they extend way beyond the small 
$z$ region where $H(t_0)d_L/c=z+(1-q(t_0))z^2/2$ (to thus enable us 
to explore the deceleration parameter $q(t)$ at both low and high 
$z$), being data which have now revealed the existence of an 
apparent cosmic repulsion. In order to present their findings in as 
quantitative a way as possible, the observers compare their Hubble 
plot data against the 
expectations of the standard Eq. (\ref{2}) cosmology, and report 
their results out as an allowed phenomenological region in the 
($\Omega_M(t_0), \Omega_V(t_0)$) parameter space. While the allowed 
region is still quite large (a situation which can be greatly 
improved on with the acquisition of further data not only in the 
key $z=1$ region, but also in the currently poorly explored 
$z=1/2$ region where there are not enough data points to constrain 
fits which go from low to high $z$), nonetheless the implications 
of the resulting allowed region are still extremely significant for 
cosmology. In particular the observers 
find \cite{Garnavich1998,Perlmutter1998} that the popular  
standard paradigm flat inflationary pure matter universe 
($\rho_M(t)=A/R^3(t)$, $\Omega_M(t_0)=1$, $\Omega_V(t_0)=0$, 
$H(t_0)d_L/c=2(1+z)(1-(1+z^2)^{-1/2})$) is actually excluded by 
their data (it predicts too low values for $d_L/z$). Similarly, 
the other extreme of a pure vacuum flat universe 
($\Omega_M(t_0)=0$, $\Omega_V(t_0)=1$, $q(t)=-1$,  
$H(t_0)d_L/c=z(1+z)$) is equally excluded (it predicts too high
values for $d_L/z$) with the data thus lying somewhere in between.
As to the allowed region itself, it is found to exhibit two
characteristic aspects. Flat universe fits can be found in 
which the cosmological constant $\Omega_V(t_0)$ term dominates over 
$\Omega_M(t_0)$ according to $\Omega_V(t_0)=3\Omega_M(t_0)$ or 
so - a situation in which the current value of the deceleration 
parameter is then found to actually be negative ($q(t_0)=-5/8$ 
according to Eq. (\ref{3})), albeit with big systematic errors 
which permit $q(t_0)$ values up to $q(t_0)=0$. Good fitting can 
also be found in the $k<0$ region. However, now, not only are there 
low mass density fits where the cosmological constant again 
dominates (according to $\Omega_V(t_0)=\Omega_M(t_0)+0.5$ or so), 
but also now (within current systematic errors) there are fits 
where $\Omega_M(t_0)$ and $\Omega_V(t_0)$ are both non-dominant, 
i.e. fits which are dominated purely by negative curvature ($k<0$ 
fits with $\Omega_M(t_0)=0$, $\Omega_V(t_0)=0$ have $q(t)=0$ and 
$H(t_0)d_L/c=(1+z)log(1+z)$). The available data are thus currently 
compatible with either of the two sources of cosmic repulsion 
($k<0$ or $\Omega_V(t_0)>0$) exhibited in Eq. (\ref{3}), and even  
appear to be not too far away from the negative curvature 
dominated $q(t_0)=0$ case, with it having to be left to future data 
to determine which, if either, of these two options is to 
ultimately be preferred. 

While we have discussed the data in terms of the Einstein-Friedmann 
evolution equation of Eq. (\ref{2}), the essence of the fits is 
that certain functional dependences of $d_L$ on $z$ and certain 
values of $q(t)$ turn out to be preferred by the data independent 
of how those values were initially derived (i.e. a specification 
of a value for $k$ and of a time dependence for $R(t)$ 
then supplies a functional dependence of $d_L$ on $z$). 
The data thus like $q(t)$ either zero or slightly negative with
the favored $d_L$ versus $z$ curve lying about midway between the 
$H(t_0)d_L/c=2(1+z)(1-(1+z^2)^{-1/2})$ and $H(t_0)d_L/c=z(1+z)$ 
curves, and being not too far away from the $k<0$ dominated 
$H(t_0)d_L/c=(1+z)log(1+z)$. 
These then are the general implications of the current data, with 
it thus being of great interest to see how specific cosmological 
models fare in meeting these constraints.
      
Now while we just noted that the standard model of Eq. (\ref{2}) is 
phenomenologically able to meet the constraints of the 
new high $z$ data since it does possess an allowed region in the   
($\Omega_M(t_0)$, $\Omega_V(t_0)$) parameter space, nevertheless, 
the  needed permitted values for these parameters create new and 
particularly severe theoretical problems for it. Specifically, no 
matter what the value of the spatial curvature $k$, the evolution 
equation of Eq. (\ref{2}) has the property that in the early 
universe the quantity $\Omega_M(t)+\Omega_V(t)$ is very close to 
one, to thus make it difficult to explain why it should be anywhere 
close to one 10 billion years or so later, with it only being able 
to be close to one today if it had somehow been fine tuned to be 
overwhelmingly close to one (to one part in $10^{60}$ or so) in the 
early universe. A candidate solution \cite{Guth1981} to this 
fine-tuning flatness problem is the existence of an inflationary 
era even prior to the RW era, a de Sitter era (actually one 
coordinate patch thereof) which would then lead to a subsequent RW 
era with $k=0$. Now as long as there was no cosmological constant 
to deal with (i.e. as long as $\Omega_V(t)$ was taken to be zero), 
inflation would then automatically fix the early universe 
$\Omega_M(t)$ to be equal to one to one part in $10^{60}$, to thus 
naturally make it of order one today. However, once there is a 
cosmological constant contribution, inflation then only fixes the 
sum of $\Omega_M(t)$ and $\Omega_V(t)$ to be close to one, and is 
not able to fix the ratio $\Omega_V(t)/\Omega_M(t)$ of these two 
contributions as well. In and of itself this would not be so severe 
an issue where it not for the fact that the new supernovae data are
now requiring $\Omega_V(t_0)$ to be of order one today. However, 
since $\Omega_M(t)$ and $\Omega_V(t)$ have radically different 
dependences on the expansion radius $R(t)$, the closeness of 
$\Omega_V(t_0)$ to one today means that it had to be zero to one 
part in $10^{60}$ in the early universe, so that the standard 
cosmology needs to be fine-tuned all over again. Since the standard 
cosmology does not fix the ratio $\Omega_V(t)/\Omega_M(t)$ in its 
current formulation it appears to be lacking a totally central 
ingredient. 

Now while the cosmological constant problem itself is of course not 
a new one, it is important to stress that the longstanding 
cosmological constant problem which predated the new supernovae 
data was actually a problem for microscopic physics as it is 
related to quantum gravity and elementary particle physics. What is 
new here is that there is now a cosmological constant problem 
associated purely with macroscopic classical cosmology, i.e. with 
the classical Einstein-Friedmann evolution equations themselves, 
and it is not at all guaranteed that a solution to the quantum 
gravity cosmological constant problem would necessarily solve this 
classical one as well. Indeed, it has been widely believed for a 
very long time now that whatever was the magic mechanism which 
would bring the cosmological constant down from 
Planck density would be one which would make $\Omega_V(t_0)$ vanish 
altogether. Even while no evidence was 
ever provided to support this viewpoint, the position was in a 
sense a (relatively) comfortable one for the community to hold in 
the past since prior observational cosmological data only required 
that $\Omega_V(t_0)$ not be larger than order $\Omega_M(t_0)$, 
something readily achievable if this unknown magic mechanism did in 
fact bring $\Omega_V(t_0)$ all the way down to zero. However, with 
these new supernovae data, this magic mechanism now has to actually 
yield an explicitly non-zero value for $\Omega_V(t_0)$, to thus 
require a one part in $10^{120}$ or so fine tuning. The current 
formulation of quantum gravity thus also appears to be lacking a 
totally central ingredient, one which would have to resolve a 
fine-tuning problem no less than $10^{60}$ or so orders 
of magnitude more severe than the one which standard classical 
cosmology now appears to have acquired.\footnote{In passing we also 
note that there is yet another potential fine tuning problem for 
$k=0$,  $\Omega_V(t_0)\neq 0$ cosmologies, namely that the relative 
contributions (i.e. the ratios) of all the various conjectured dark 
matter candidates needed for $\Omega_M(t_0)$ would need to all be 
jointly tuned so that they would then collectively sum to precisely 
$1-\Omega_V(t_0)$. And, if indeed, this is to be achieved without 
fine tuning it would appear that  
these same dark matter particles (rather than some only remotely 
related scalar inflaton field) would have to play an explicit and 
direct dynamical role in collectively fixing the early universe 
topology in the first place.} Now of course there is a lot of 
physics in quantum cosmology and string theory which has yet to be
understood, and the unification of Einstein gravity with 
quantum theory may yet resolve this issue to naturally lead to an 
early universe cosmology with an evolution which possibly even 
takes one of the forms (perhaps one with $k<0$) exhibited in the 
solutions of Eqs. (\ref{10}) and (\ref{11}). However, it is now 
extremely urgent for these theories, if they are to be theories of 
gravity, to explain just exactly why $\Omega_V(t_0)$ is in fact of 
order one today, and in a sense their very relevance to the real 
world rests on their achieving precisely that.

As we see, cosmology has great trouble trying to respect the 
constraints imposed by the standard model evolution equation of 
Eq. (\ref{2}), and, indeed, all of the problems of the standard 
cosmology can be traced 
\cite{Mannheim1990,Mannheim1995,Mannheim1998} to one single source, 
namely the a priori use of Eq. (\ref{2}) in the first place, with 
Eq. (\ref{2}) simply not appearing to be robust enough an 
evolution equation to readily accommodate the constantly changing 
requirements of data. Now, even though the community would only be 
prepared to adopt any possible alternative to standard Einstein 
gravity with the greatest of reluctance, nonetheless the very 
severity of the situation described above does warrant that it at 
least consider such a possibility. Moreover, even beyond 
phenomenological issues, the central theoretical weakness of 
standard gravity is \cite{Mannheim1994} that within the class of 
covariant, pure metric based theories of gravity, there 
is currently no known fundamental principle which would actually 
uniquely select out the Einstein-Hilbert action from amongst the 
infinite class of equally covariant metric theories  
based on higher order derivative actions that could just as readily 
be considered. And indeed, it is the very absence of any 
such principle which has engendered problems such as the 
cosmological constant problem in the first place, with the 
covariance principle itself simply not being sufficiently 
restrictive to unambiguously fix the gravitational equations of 
motion. Standard gravity thus appears to currently lack a 
fundamental principle, and indeed, it is none other than this 
very shortcoming which actually permits the exploration of 
candidate alternatives. However, since any potential candidate 
higher order derivative alternative theory of gravity would 
immediately be expected to yield essentially non-predictive higher 
order derivative cosmological evolution equations as well, we must 
thus look for a pure metric based 
alternate theory which (i) possesses some fundamental 
principle which does unambiguously fix the gravitational action, 
(ii) which naturally controls the cosmological constant, (iii) 
which still has a low order evolution equation with as much 
predictive power as the standard one, and (iv) which naturally 
incorporates either or both of the two cosmic repulsion mechanisms 
we identified above. Quite remarkably there actually does exist 
such a theory, namely conformal gravity, a gravitational theory 
which is based on an underlying local   
$g_{\mu \nu} (x) \rightarrow \Omega (x) g_{\mu \nu} (x)$ 
conformal invariance, an invariance which then yields 
$I_W=-\alpha \int d^4x (-g)^{1/2} C_{\lambda\mu\nu\kappa} 
C^{\lambda\mu\nu\kappa}$ (where $ C_{\lambda\mu\nu\kappa}$ is 
the conformal Weyl tensor and $\alpha$ is purely dimensionless) 
as its unique gravitational action. Such a theory has been shown 
(i) capable of keeping the cosmological constant under control 
\cite{Mannheim1990,Mannheim1995,Mannheim1998} (viz. a fundamental 
cosmological constant is forbidden by the underlying conformal 
invariance, while the magnitude of any induced one is 
naturally constrained by the tracelessness of the conformal 
energy-momentum tensor to be within acceptable observational 
limits), (ii) to have no flatness (or horizon) problem and to 
naturally yield a $k<0$ universe 
\cite{Mannheim1992,Mannheim1995,Mannheim1997,Mannheim1996}, 
(iii) to thereby 
resolve the cosmic repulsion problem \cite{Mannheim1998}, and (iv) 
to still contain \cite{MannheimandKazanas1989} the Schwarzschild 
solution to thus still recover the standard relativistic solar 
system predictions despite its not being based on the 
Einstein-Hilbert action. Moreover, despite not being based on the 
Einstein-Hilbert action, in the conformal theory an effective 
Einstein-Hilbert term is nonetheless induced cosmologically 
\cite{Mannheim1990,Mannheim1992}, with conformal cosmology  
being found to actually be described by a second order equation of 
motion, viz.
\begin{equation}
U^\mu U^\nu(\rho_{M}+p_{M})/c+g^{\mu\nu}p_{M}/c
-\hbar S^2(R^{\mu\nu}-
g^{\mu\nu}R^\alpha_{\phantom{\alpha}\alpha}/2)/6            
-g^{\mu\nu}\hbar\lambda S^4=0
\label{13}
\end{equation}
where $S$ is the expectation value of a scalar field which 
serves to spontaneously break the conformal symmetry. While the 
non-vanishing of $S$ thus induces an effective Einstein tensor into
the dynamics, it is important to note that this term is induced 
with a negative rather than a positive coefficient, something that 
will turn out to have major consequences for the predictions of the
theory. With the coefficient of this induced Einstein tensor then 
depending on the magnitude of $S$, and with the actual magnitude 
of this expectation value itself then being self-consistently
determined in Eq. (\ref{13}) by however many particle states are
occupied in the energy density $\rho_{M}$, the resulting induced 
cosmological constant term $\hbar\lambda S^4$ is thus constrained 
to be of the same of order of magnitude as all the other terms in 
Eq. (\ref{13}). It is in this way then that the cosmological 
constant is kept under control in conformal gravity. 

The evolution equations which are to 
replace Eqs. (\ref{2}) and (\ref{3}) take the remarkably similar 
forms 
\begin{equation}
\dot{R}^2(t) +kc^2 =
-3\dot{R}^2(t)(\Omega_{M}(t)+
\Omega_{V}(t))/ 4 \pi S^2 L_{PL}^2
\equiv \dot{R}^2(t)\bar{\Omega}_{M}(t)+
\dot{R}^2(t)\bar{\Omega}_{V}(t)
\label{14}
\end{equation}
and
\begin{equation}
q(t)=(n/2-1)\bar{\Omega}_{M}(t)-\bar{\Omega}_{V}(t)=
(n/2-1)(1+kc^2/\dot{R}^2(t)) - n\bar{\Omega}_{V}(t)/2 
\label{15}
\end{equation}
and thus only contain familiar ingredients. (Here the parameter
$\Lambda$ in the definition $\Omega_{V}(t)=
8\pi G\Lambda/3cH^2(t)$ is given by $\hbar\lambda S^4$). Indeed, 
as we see, Eq. (\ref{14}) only differs from the analogous standard 
model Eq. (\ref{2}) in one regard, namely that the 
quantity $-\hbar S^2 /12$ has replaced the familiar 
$c^3/16 \pi G$. Given this one simple change in sign, 
the conformal theory is then found to behave very 
differently from the standard theory. Specifically, in the 
(simpler to treat) radiation era where $\rho_M=A/R^4 =\sigma T^4$, 
Eq. (\ref{14}) may be rewritten as
\begin{equation}
R^2(t)\dot{R}^2(t)-\alpha c^2(R^2(t)-R^2_{+})
(R^2(t)-R^2_{-})=0
\label{15a}
\end{equation}
where we have introduced the parameters 
$\alpha =-2\lambda S^2$, 
$\beta =(1+8A\alpha/k^2\hbar c S^2)^{1/2}$, 
$R^2_{\pm}=k(1\pm\beta)/2\alpha$). As we see, precisely 
because of this sign change (a sign change which makes the 
conformal theory act like a standard model with repulsive rather 
than attractive gravity), all conformal cosmological 
solutions have the property that $R(t)$ does not vanish at the 
initial time $t=0$ (defined as the time where $\dot{R}(t)$ is 
zero), so that all of the cosmologies are singularity free and 
have non-zero minimum radius. Consequently the early 
universe $\Omega_{M}(t=0)$ is automatically infinite (i.e. nowhere 
close to one), with no possible conformal cosmology then having a 
flatness problem \cite{Mannheim1992,Mannheim1995}. Thus, unlike 
the standard gravity case, in conformal gravity all the $k\neq 0$ 
cosmologies are natural in the sense that they require no fine 
tuning. At $t=0$ Eq. (\ref{15a}) 
possesses two branches of solutions ($R^{2}_{+}$ and 
$R^{2}_{-}$), which are selected according to the 
sign of $k$. And indeed, since the effect of the matter term 
$\bar{\Omega}_{M}(t)$ in both the $\lambda > 0$ and 
$\lambda =0$ cases in Eq. (\ref{14})  
is to produce a scalar curvature $k$ which can 
only be negative (to thus immediately yield the cosmic repulsion 
which had previously been found to naturally occur in the conformal 
cosmology study of Ref. \cite{Mannheim1998}), negative $k$ 
solutions can then be found for $\lambda$ positive or zero, and
in fact even for $\lambda$
negative (as long as $\bar{\Omega}_{V}(t)$ does not overwhelm 
$\bar{\Omega}_{M}(t)$), with the complete set 
of solutions associated with the $k<0$, $R^{2}_{-}$ branch being 
given by 
\begin{eqnarray}
R^2(t,\alpha<0,k<0)=k(1-\beta)/2\alpha+
k\beta sin^2 ((-\alpha)^{1/2} ct)/\alpha
\nonumber \\
R^2(t,\alpha=0,k<0)=-2A/k\hbar c S^2-kc^2t^2
\nonumber \\
R^2(t,\alpha>0,k<0)= -k(\beta-1)/2\alpha
-k\beta sinh^2 (\alpha^{1/2} ct)/\alpha
\label{16}
\end{eqnarray}
\noindent
The  $R^{2}_{+}$ branch can only be realized when $\lambda<0$ and 
$k \geq 0$, to then give solutions of the form 
\begin{eqnarray}
R^2(t,\alpha > 0,k=0)=(-A/\hbar\lambda c S^4)^{1/2}
cosh(2\alpha^{1/2}ct)
\nonumber \\
R^2(t,\alpha > 0,k>0)=k(1-\beta)/2\alpha+
k\beta cosh^2 (\alpha^{1/2} ct)/\alpha
\label{17}
\end{eqnarray}
\noindent
For these various solutions the associated deceleration parameters 
are readily calculated, and they take the respective forms
\begin{eqnarray}
q(t,\alpha<0,k<0)=
(1+cos^2(2(-\alpha)^{1/2}ct)-2cos(2(-\alpha)^{1/2}ct)/\beta)/
sin^2(2(-\alpha)^{1/2}ct)
\nonumber \\
q(t,\alpha=0,k<0)=-2A/k^2\hbar c^3 S^2t^2
\nonumber \\
q(t,\alpha>0,k<0)=
(-1-cosh^2(2\alpha^{1/2}ct)+2cosh(2\alpha^{1/2}ct)/\beta)/
sinh^2(2\alpha^{1/2}ct)
\nonumber \\
q(t,\alpha>0,k=0)=-1-2/sinh^2(2\alpha^{1/2}ct)
\nonumber \\
q(t,\alpha>0,k>0)=
(-1-cosh^2(2\alpha^{1/2}ct)-2cosh(2\alpha^{1/2}ct)/\beta)/
sinh^2(2\alpha^{1/2}ct)
\label{17a}
\end{eqnarray}

While the above solutions apply in all epochs, it turns out that 
they simplify a great deal in the current era. Specifically, since 
all of these solutions have a non-zero minimum radius, they all 
have some very large but finite maximum temperature $T_{max}$. 
Explicit calculation in the illustrative $\lambda=0$, $k<0$ case 
then yields \cite{Mannheim1995} the temperature dependence
\begin{equation}
T^2_{max}/T^2(t)=1/ (1-tH(t))=1-1/ q(t)
=1+4\pi S^2 L^2_{PL}/ 3\Omega_{M}(t)
=1-1/\bar{\Omega}_{M}(t)
\label{18}
\end{equation}
to entail that the scale parameter $S$ of the model must be many 
orders of magnitude larger than $L^{-1}_{PL}$ (in consequence of
which the current value of the deceleration parameter would then be 
zero and the current age of the universe would then be given by
the phenomenologically acceptable value of $1/H(t_0)$ in this 
$\lambda=0$ cosmology). With this condition on $SL_{PL}$ being 
found to also hold for all the other solutions in Eqs. (\ref{16}) 
and (\ref{17}) as well, and, moreover, to even carry over to 
the matter $\rho_M=A/R^3$ era \cite{Mannheim1995}, we see 
that in conformal cosmology the matter contribution is completely 
suppressed at current temperatures (it thus only dominates in the 
early universe where it sets the scale for the minimum radius in 
Eqs. (\ref{16}) and (\ref{17})), with the current era Eq. 
(\ref{14}) then reducing to
\begin{equation}
\dot{R}^2(t) +kc^2 =
\dot{R}^2(t)\bar{\Omega}_{V}(t)
\label{19}
\end{equation}
\noindent
Comparison with Eq. (\ref{2}) shows that current era conformal 
cosmology thus looks exactly like a low mass standard model 
cosmology, except that instead of $\Omega_{M}(t_0)$ being small 
(something difficult to understand in the standard theory) it is 
$\bar{\Omega}_{M}(t_0)=-3\Omega_{M}(t_0)/4\pi S^2 L^2_{PL}$ which 
needs to be small instead, a thus
completely natural expectation of conformal cosmology which is
completely compatible with the Eq. (\ref{2}) based low 
$\Omega_{M}(t_0)$ fitting found for the new supernovae data.

Additionally, the current era Eq. (\ref{19}) is completely 
analogous to the purely geometrical Eq. (\ref{9}), with the 
solutions of Eqs. (\ref{16}) and (\ref{17}) then reducing to those 
given in  Eqs. (\ref{10}) and (\ref{11}) in the $A \rightarrow 0$ 
limit.\footnote{Except for the $\alpha>0$, $k=0$ case where the 
difference between the two $R(t,\alpha>0,k=0)$ solutions given in 
Eqs. (\ref{10}) and (\ref{17}) is due to the fact that in the pure 
$A=0$ de Sitter case with no matter $\dot{R}(t)$ has to vanish
at $t=-\infty$ rather than at $t=0$.} The discussion of the 
deceleration parameter then essentially parallels that of 
Eq. (\ref{12}) to thus automatically recover all of its 
accelerating universe solutions, with conformal gravity 
currently thus being able to naturally fit the available Hubble 
plot data (the data currently permit a range of values for 
$\bar{\Omega}_{V}(t_0)=-3\Omega_{V}(t_0)/4\pi S^2 L^2_{PL}$ 
(a range which includes zero), and are thus not yet able to pin 
down an explicit value for $\lambda$). Depending on 
whether the parameter $\lambda$ is zero or not 
conformal cosmology thus naturally incorporates either or both of 
the two kinematic cosmic repulsion mechanisms ($k<0$ and 
$\alpha\neq 0$) we identified earlier in this paper, with it thus 
being of great interest to see which part, if any, of the conformal 
gravity $(k,~\lambda)$ parameter space would be able to meet 
future data. 

As regards the conformal cosmology solutions, we see that one of 
them, $R(t,\alpha<0,k<0)$, leads to a recontracting 
universe which actually accelerates for part of the time. Indeed, 
while all the other four solutions given in Eqs. (\ref{16}) and 
(\ref{17}) possess deceleration parameters which are always less 
than or equal to zero (and which approach the corresponding values 
given in Eq. (\ref{12}) at late times), $q(t,\alpha<0,k<0)$ 
actually changes sign during its evolution. Specifically, for this 
particular cosmology $R^2(t)$ goes from its minimum value of 
$k(1-\beta)/2\alpha$ to its maximum value of $k(1+\beta)/2\alpha$ 
in a time given by $\tau/2=\pi/2(-\alpha)^{1/2}c$, with a full 
cycle thus taking a time $\tau$. However, while $\dot{R}(t)$ only 
changes sign at $t=\tau/2$, $\ddot{R}(t)$ changes sign when 
$cos(2\pi t/\tau)=(1-(1-\beta^2)^{1/2})/\beta $. Since 
$\ddot{R}(t)$ is positive at $t=0$ and negative at $t=\tau/2$, it 
thus changes sign somewhere in between, even while $\dot{R}(t)$ 
does not undergo any analogous such sign change. Hence the fact 
that $\ddot{R}(t)$ starts off accelerating does not imply that the 
cosmology will continue to accelerate indefinitely. In and of 
itself then the detection of an acceleration in some particular 
epoch is thus not an indicator that the universe will 
necessarily continue to
expand for ever.\footnote{As we thus see, it is not sufficient to 
simply measure current era cosmological parameters such as the 
current value of the mass density in order to ascertain the 
universe's ultimate fate. Rather, in order to predict the future
evolution of the universe, it is necessary to also know which 
particular cosmological evolution equation is the appropriate one 
for the future, with Eqs. (\ref{2}) and (\ref{14}) giving very 
different futures (and histories) even with the same current era 
value for $\Omega_{M}(t_0)$.} 

As a cosmology, Eq. (\ref{14}) 
exhibits three distinct epochs, the earliest being particle physics 
($\rho_{M}$) dominated (an era where $k$ is fixed once and for 
all), the intermediate being curvature dominated, and the final one 
being $\lambda$ dominated. The scalar fields which initially gave 
mass scales to elementary particle physics and to gravity way back 
in the early universe will thus ultimately come to dominate the 
expansion rate $R(t)$, with the geometries of Eqs. (\ref{16}) and 
(\ref{17}) evolving at late times into the de Sitter and anti de 
Sitter solutions of Eqs. (\ref{10}) and (\ref{11}), something which 
is actually geometrically natural (both locally and globally) in a 
theory with an underlying conformal invariant structure since both 
$O(4,1)$ and $O(3,2)$ are subgroups of the full $O(4,2)$ conformal 
group.\footnote{In passing we note that recent studies in M/string 
theory \cite{Maldacena1997} have also uncovered a possible 
connection between conformal field theories and anti de Sitter 
spacetimes. It might thus be of some interest to see whether those 
studies bear any relation to the connection between conformal 
gravity and anti de Sitter geometry found in the present paper.} In 
conclusion we thus see that conformal cosmology appears to have the 
capability to address many pressing cosmological problems, and that
it provides a very different framework in which to view them. Given 
the fact that the recently detected cosmic repulsion is putting 
such severe pressure on the standard model, conformal gravity 
would thus appear to merit further exploration. This work has been 
supported in part by the Department of Energy under grant No. 
DE-FG02-92ER40716.00.

\end{document}